# Room Temperature Giant Charge-to-Spin Conversion at SrTiO$_3$/LaAlO$_3$ Oxide Interface

Yi Wang,[†,§] Rajagopalan Ramaswamy,[†,§] M. Motapothula,[†] Kulothungasagaran Narayanapillai,[†] Dapeng Zhu,[†] Jiawei Yu,[†] T. Venkatesan,[†,‡] and Hyunsoo Yang[*,†]

[†]Department of Electrical and Computer Engineering, NUSNNI, National University of Singapore, 117576, Singapore

[‡]Department of Physics, National University of Singapore, Singapore, 117542, Singapore

**ABSTRACT:** Two-dimensional electron gas (2DEG) formed at the interface between SrTiO$_3$ (STO) and LaAlO$_3$ (LAO) insulating layer is supposed to possess strong Rashba spin-orbit coupling. To date, the inverse Edelstein effect (i.e. spin-to-charge conversion) in the 2DEG layer is reported. However, the direct effect of *charge-to-spin* conversion, an essential ingredient for spintronic devices in a current induced spin-orbit torque scheme, has not been demonstrated yet. Here we show, for the first time, a highly efficient spin generation with the efficiency of ~6.3 in the STO/LAO/CoFeB structure at room temperature by using spin torque ferromagnetic resonance. In addition, we suggest that the spin transmission through the LAO layer at high temperature range is attributed to the inelastic tunneling via localized states in the LAO band gap. Our findings may lead to potential applications in the oxide insulator based spintronic devices.

[*]E-mail: eleyang@nus.edu.sg

[§]Y.W. and R.R. contributed equally to this work.

The SrTiO$_3$ (STO) and LaAlO$_3$ (LAO) are two wide-band-gap insulators, however, when both are interfaced each other, a conductive two-dimensional electron gas (2DEG) is formed at their interface.[1] This 2DEG possesses exotic properties, such as superconductivity and ferromagnetism,[2-4] which makes the STO/LAO heterostructure a unique system. Recently, a strong Rashba spin-orbit coupling (SOC) has been theoretically reported in the *d* bands of the 2DEG at the STO/LAO interface,[5-7] leading to the orbital and spin degrees of freedom to be strongly coupled with each other. Therefore, the STO/LAO heterostructure can enable potential spintronic innovations and now is attracting rising research interests.[8-13] Experimentally, Narayanapillai *et al.*[8] have reported an extremely strong charge current induced Rashba field in the 2DEG layer, which verifies the presence of strong Rashba SOC predicted at the STO/LAO interface. Most recently, the inverse Edelstein effect at the STO/LAO interface has been demonstrated by using the spin pumping technique,[10-12] which shows a notable spin-to-charge conversion. Additionally, a long spin diffusion length over 300 nm in the 2DEG channel has also been reported.[9,13] These observations suggest that the STO/LAO heterostructure can be a good spin detector as well as spin channel, which is promising to advance the realization of long sought spin transistors. However, the direct *charge-to-spin* conversion at the STO/LAO interface has not been demonstrated yet. Moreover, the identification of highly efficient spin current generation is an essential step for the modern magnetic memory in a spin-orbit torque (SOT) driven magnetization switching scheme, as reported in heavy metals, such as Pt, Ta and W, recently.[14-17]

In this study, we experimentally show a giant room temperature charge-to-spin conversion efficiency (i.e. SOT efficiency) in the STO/LAO/Co$_{15}$Fe$_{60}$B$_{25}$ structure by the spin torque ferromagnetic resonance (ST-FMR) technique. The SOT efficiency ($\theta_\parallel$) is evaluated to be ~6.3 at room temperature, which is almost two orders of magnitude larger than the spin Hall angles in heavy metals, such as Pt and Ta.[15,18,19] From the temperature dependent ST-FMR measurements, we have clarified that the mechanism of inelastic tunneling via localized states in the LAO band gap accounts for the spin transmission through the LAO layer. Our demonstration provides a new platform for the innovations of SOT based spintronic applications utilizing oxide interfaces.

A layer of 8 unit cells (uc) LAO (1 uc ≈ 0.39 nm) is grown on a $TiO_2$ terminated atomically smooth single crystalline STO (001) substrate by pulsed laser deposition (PLD) at 750 °C in an oxygen partial pressure of 1 mTorr, which was pre-treated with buffered oxide etch and air-annealed at 950 °C for 1.5 hour. The growth is monitored by in-situ reflective high energy electron diffraction (RHEED). As shown in Figure 1a, the RHEED oscillations indicate that the LAO growth proceeds layer by layer. A standard Hall bar is fabricated for four-probe electrical measurements (Supporting Information S1). As shown in Figure 1b, the sheet resistance $R_S$ decreases drastically as the temperature decreases and then almost saturates at ~10 K. This temperature dependence implies a typical metallic behavior[1,10] of the 2DEG at the STO/LAO interface (denoted in the inset of Figure 1b). The sheet carrier concentration $n_{2D}$ is estimated to be ~$1.17×10^{14}/cm^2$ at room temperature. The properties of the 2DEG layer are consistent with previous reports.[20,21]

The ST-FMR measurement is an effective technique to evaluate the SOT efficiency as shown in previous works.[18,22] Figure 2a shows the schematic diagram of ST-FMR measurement setup and the structure of ST-FMR device, which has the structure of STO substrate (001)/LAO (8 uc, ~3.1 nm)/$Co_{15}Fe_{60}B_{25}$ (CFB, 7 nm)/MgO (2 nm)/$SiO_2$ (4 nm). In the ST-FMR device, a 7-nm thick CFB film with MgO/$SiO_2$ capping is grown by dc magnetron sputtering at room temperature. The ST-FMR device fabrication process is described in the Supporting Information S1. The left panel of Figure 2a illustrates the current induced SOTs and the SOT driven magnetization dynamics in STO/LAO/CFB structures. As an in-plane radio-frequency (rf) current ($I_{RF}$) is applied in the STO/LAO 2DEG layer, non-equilibrium spins are generated due to the Rashba-Edelstein effect.[10-12,23,24] Subsequently, these spins are absorbed by the CFB layer and exert oscillating damping-like torque ($\tau_{DL}$) and/or a field-like torque ($\tau_{FL}$) on the CFB local magnetization. In addition, the rf current induces Oersted field ($H_{RF}$) torque ($\tau_{Oe}$). All these torques make the CFB magnetization into precession with the same frequency as $I_{RF}$. Consequently, the product of the oscillation of the anisotropic magnetoresistance ($\Delta R$) and the $I_{RF}$ produce a mixing dc voltage $V_{mix}$ (i.e. ST-FMR signal) across the device,[18,19,22,25] which is detected by a lock-in amplifier. An in-plane external magnetic field ($H$) is applied at a fixed angle ($\theta_H$) of 38º with respect to $I_{RF}$. The measurements are performed with a nominal rf power of 15 dBm at room temperature.

Figure 2b shows the resonance $f$ as a function of resonance magnetic field $H_0$ for a typical ST-FMR device, which can be fitted with the Kittel formula $f = (\gamma/2\pi)[H_0(H_0 + 4\pi M_{eff})]^{1/2}$, where $\gamma$ is the gyromagnetic ratio. From fitting, the effective magnetization ($4\pi M_{eff}$) for the CFB layer is determined to be ~1.18 T. Figure 2c shows one representative ST-FMR signal $V_{mix}$ (open symbols) at 6 GHz, which is fitted by $V_{mix} = V_S F_S + V_A F_A$, where $F_S$ and $F_A$ are the symmetric and antisymmetric Lorentzian functions, respectively. The amplitude of symmetric ($V_S$) and antisymmetric component ($V_A$) are correlated with $\tau_{DL}$ and ($\tau_{FL} + \tau_{Oe}$), respectively.[18,22] A large symmetric component $V_S F_S$ is observed, suggesting a large in-plane spin polarization and thus $\tau_{DL}$. Moreover, as shown in the inset of Figure 2c, the symmetric component $V_S F_S$ is negligible in a control ST-FMR device having the structure of Si/SiO$_2$ sub/CFB (7 nm)/MgO (2 nm)/SiO$_2$ (4 nm) with a same dimension as the device in Figure 2c. This verifies that the main contribution of the in-plane spin polarization and the induced $\tau_{DL}$ to the CFB magnetization dynamics is from the STO/LAO interface. Further, we also find that the magnetization precession of our device under the rf power of 15 dBm is in the linear regime (Supporting Information S2, Figure S1).

The SOT efficiency $\theta_\parallel$ (= $J_S/J_C$) in the STO/LAO/CFB systems can be evaluated from only the fitted $V_S$ by utilizing the established method[22,25] (also see Supporting Information S3 and S4, Figure S2 and Table S1), where $J_S$ represents the spin current density flowing into the CFB layer and $J_C$ (A/cm$^2$) is the uniform charge current density in the STO/LAO 2DEG layer. Subsequently, we obtain a large $\theta_\parallel$ ~6.3 ± 1 at room temperature arising from the interface between the STO and LAO layer. This value is one or two orders of magnitude higher than that of the heavy metals, such as Pt, Ta and W, having spin Hall angle of ~0.06–0.3.[15,16,18,19] Moreover, the performance of spin current generation in the STO/LAO heterostructure is similar or even better than the emerging topological insulator Bi$_2$Se$_3$ with SOT efficiency of ~0.43–3.5.[22,25,26] Our observation further confirms the previous report, which shows a strong current induced Rashba effective field $H_R$ with the value of ~1.76 T for a charge current density of $J_C$ = 10$^5$A/cm$^2$.[8]

In our STO/LAO/CFB device, there is a ~3.1-nm thin LAO insulting layer separating the STO/LAO spin source and CFB magnetic layer, which is different from previous cases with a direct contact of the heavy metal (or topological insulator) with a ferromagnetic layer. However, our observation of a large SOT efficiency at room temperature indicates that the a large amount of spins can transmit through the LAO insulating layer and are absorbed by the CFB magnetization. There are two possible mechanisms which can account for this spin transmission. The first mechanism is the spin dependent direct elastic tunneling observed in extensively studied magnetic tunnel junctions. For this scenario, the barrier is typically very thin with the thickness ~2 nm.[27-30] However, the LAO layer in our device is relatively thicker, at ~3.1 nm, leading to a lesser overlap of the electron wave functions from the two sides of LAO layer, and thus, a low direct elastic tunneling amplitude. Therefore, we expect that the spin dependent direct elastic tunneling in the LAO layer is not the main mechanism for the spin transmission in our samples, which is confirmed by our temperature dependent ST-FMR measurements, as discussed later.

The second possible mechanism is the inelastic tunneling (IET) process which can be predominant when the tunneling barrier becomes thicker.[31-33] It is known that the defects, such as the oxygen vacancies ($V_O$), can be present in the LAO layer during film growth and form localized states (LS) in the LAO band gap.[34-37] Hence, the IET process via these LS[31-33,38-40] might account for the spin transmission in our devices. The schematic of the IET process is depicted in the inset of Figure 3a, where the spin-polarized electrons (denoted by the blue dots) in the 2DEG layer inelastically tunnel into the CFB layer via the LS (the oxygen vacancies $V_O$ are denoted by the short red lines) in the LAO band gap with the assistance of phonon or electrical field. If the IET via LS is the main mechanism of the spin transmission through the LAO layer, the SOT efficiency is expected to exhibit a very strong temperature dependence as indicated by the Glazman-Matveev (GM) theory,[31,33,38,41], which describes the IET with a power law dependence on temperature.

In order to confirm the above speculation and understand the underlying mechanism, we have performed the ST-FMR measurements at different temperatures. Figure 3a shows $\theta_\parallel$ as a function of temeprature. We find that $\theta_\parallel$ exhibits the largest value (~6.3) at room temperature and then decreases rapidly as the temperature decreases to ~150 K. Finally, $\theta_\parallel$ becomes negligible around 50 K within our measurement resolution. The significant temperature dependence of $\theta_\parallel$ suggests that the conduction of spin polarized electrons through the LAO layer decreases drastically as temperature decreses to the low temperature range. This cofirms that the mechanism of spin dependent direct elastic tunneling in the LAO layer can not account for the spin transmission in our samples, which is supposed to show a weak temperature dependence.[33] Rather, the observation of the strong temperature dependent $\theta_\parallel$ is in line with the IET via LS in the LAO layer, which is the alternative possible mechanism having discussed above. It is to be noted that the current shunting in the STO/LAO 2DEG layer cannot explain the temperature dependent behavior in our ST-FMR measurements shown in Figure 3a, since the anisotropic magnetoresistance only slightly decreases by ~20% as temperature decreases from 300 to 18 K (see Supporting Information S5, Figure S3).

Further, we can write the measured SOT efficiency as $\theta_\parallel = J_S/J_C = J_{S\text{-intri}}\eta/J_C$, where $J_{S\text{-intri}}$ is the spins generated in the STO/LAO 2DEG layer, $J_{S\text{-intri}}/J_C$ is the intrinsic SOT efficiency ($\theta_{\parallel\text{-intri}}$) in this 2DEG layer and $\eta$ denotes the inelastic tunneling efficiency of spin-polarized electrons. The $\eta$ is proportional to $G/G_0$, where $G$ is the electron conductance through the LAO layer at different temperature, $G_0$ is the ideal conductance without the LAO barrier layer and is assumed to be constant. On the basis of GM model,[31,33,38,41] the temperature dependent $G$ is

$$G = c_1 + \sum_{N\geq 2}^{N} c_N T^{N-2/(N+1)}$$

, where $N$ is the number of LS in the conduction chains in the LAO layer (referred as $N$-LS chains), $T$ is the temperature, and the constant $c_1$ and $c_N$ represent the conduction contribution from direct elastic tunneling & 1-LS chains, and $N$-LS chains ($N \geq 2$), respectively. Consequently, the SOT efficiency can be rewritten as $\theta_\parallel = C_1 + \sum_{N\geq 2}^{N} C_N T^{N-2/(N+1)}$, where $C_N$ ($= c_N \theta_{\parallel\text{-intri}}/G_0$) are the free parameters for fitting. As shown in Figure 3a, we find that the temperature dependent $\theta_\parallel$ can be well fitted with the LS number $N$ up to three (red curve). It is noted that we take the upper bound and assume that the spins are conserved during the IET process of the spin-polarized electrons in the 2DEG layer following the recent reports.[31,42] We believe that this fitting can well capture the core mechanism of our results even though the spin flip might happen during the IET process. Further, the respective contributions of direct elastic tunneling & 1, 2 and 3-LS chains are also obtained from fitting, as shown in Figure 3a.

From the fit, the conduction weight of respective conduction chains can be obtained by normalizing its conductance to the whole conductance at each temperature. The results are shown in Figure 3b. We find that at low temperatures, the conductance is dominantly governed by direct elastic tunneling & 1-LS chains, however the total amount of spin transmission is negligible leading to a very small $\theta_\parallel$ as shown in Figure 3a (black curve). This is expected since less phonons and LS are available at low temperatures. As temperature increases, the chains with increasing numbers of LS come into play and the IET via 3-LS chains dominates the conductance. Therefore, we attribute that the IET of spin-polarized electron through LAO via the LS is the main mechanism for our observation at room temperature. Our above analysis may provide clues for the spin transmission mechanism in the most recently reported spin pumping measurements in the STO/LAO/NiFe systems.[10,12] Additionally, by taking the 2DEG thickness of 10 nm,[43] we can estimate the corresponding *interface* SOT efficiency,[44,45] $\lambda$ (nm$^{-1}$) to be ~0.63 nm$^{-1}$ at room temperature (Supporting Information S6). This value is similar as the recently reported interface SOT efficiency from a topological insulator $(Bi_{1-x}Sb_x)_2Te_3$.[45]

Figure 4 shows the sheet carrier concentration $n_{2D}$ of the 2DEG layer at different temperatures. We find that $n_{2D}$ decreases from ~11.7× 10$^{13}$ to 1.14 × 10$^{13}$/cm$^2$ as temperature decreases from 300 to 10 K. Especially, as the temperature decreases to the range of ~20–30 K, $n_{2D}$ becomes around ~1.8 × 10$^{13}$/cm$^2$, which indicates that the Fermi energy is around the Lifshitz point and the SOC strength in the 2DEG is maximum.[6] The inset of Figure 4 shows the sketch of the relationship between the strength of SOC and Fermi energy adapted from Ref. 6. This indicates that the spins can be effectively generated with a higher $\theta_{\parallel\text{-intri}}$ in the 2DEG layer at low temperturess. However, our ST-FMR measurements shows that $\theta_\parallel$ becomes negligible at low temperature range as shown in Figure 3a. Since the measured $\theta_\parallel$ is determined by the spin-orbit coupling in the 2DEG (i.e. $\theta_{\parallel\text{-intri}}$) as well as the spin transmission through the LAO layer, the abovementioned discrepancy is ascribed to the lesser spins transmission through the LAO layer at low temperatures. However, close to room temperature, the spin transmission increases significantly. This further corroborates our conjecture, the IET of spin-polarized electrons via LS governs the amount of spins injected into CFB and thus the SOTs exerted on the CFB magnetization. Lastly, it is worth noting that, during this spin transmission through the LAO layer, the amount of spins that actually reach the CFB is reduced due to the transmission coefficient ($< 1$) and possible spin flip in the LAO layer as well as the spin memory loss[46,47] at the STO/LAO and LAO/CFB interfaces. Hence, $\theta_\parallel$ (see Figure 3a) are the effective values and are the lower bounds of $\theta_{\parallel\text{-intri}}$ in the STO/LAO 2DEG layer (also see Supporting Information S3).

One crucial ingredient for the exploration of high-density non-volatile magnetic memories or spin transistors is the highly efficient spin generation. We directly demonstrate a giant SOT efficiency in the STO/LAO/CFB heterostructures at room temperature for the first time, which verifies the strong SOC at the STO/LAO interface and suggests that the STO/LAO can be a promising spin generator for applications. In addition, we discuss insight into the mechanism of spin transmission through the LAO layer, which has not been clear in recent STO/LAO spin pumping reports. Our results may greatly invigorate the room temperature SOT induced magnetization switching in STO/LAO/ferromagnet or ferrimagnet structures as well as the full oxide insulator based spin devices.

**ACKNOWLEDGMENTS**

This work was partially supported by the National Research Foundation (NRF), Prime Minister's Office, Singapore, under its Competitive Research Programme (CRP Award No. NRFCRP12-2013-01).

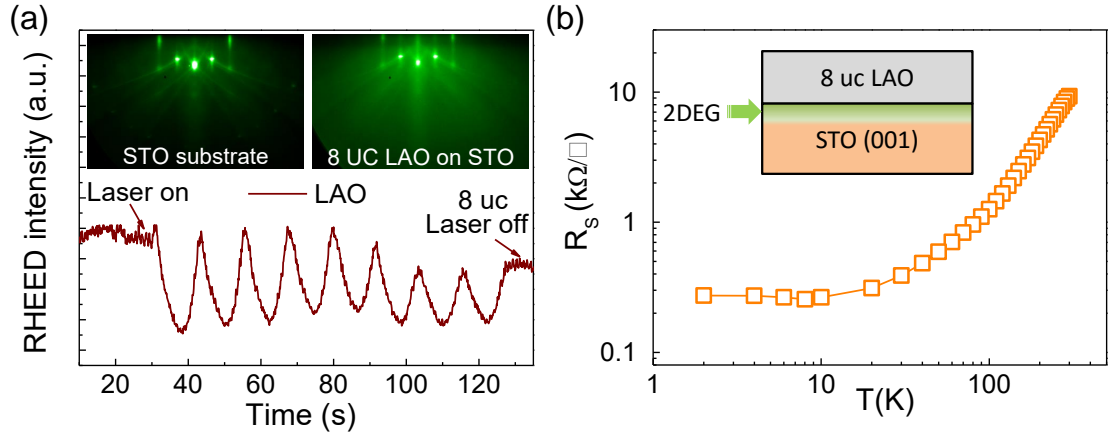

**Figure 1.** (a) RHEED oscillations during the growth of the 8-uc LAO film and RHEED diffraction patterns of STO substrate before (left inset) and after LAO layer growth (right inset). (b) The temperature dependent sheet resistance $R_S$ of STO/LAO (8 uc) heterostructures by four-probe measurements.

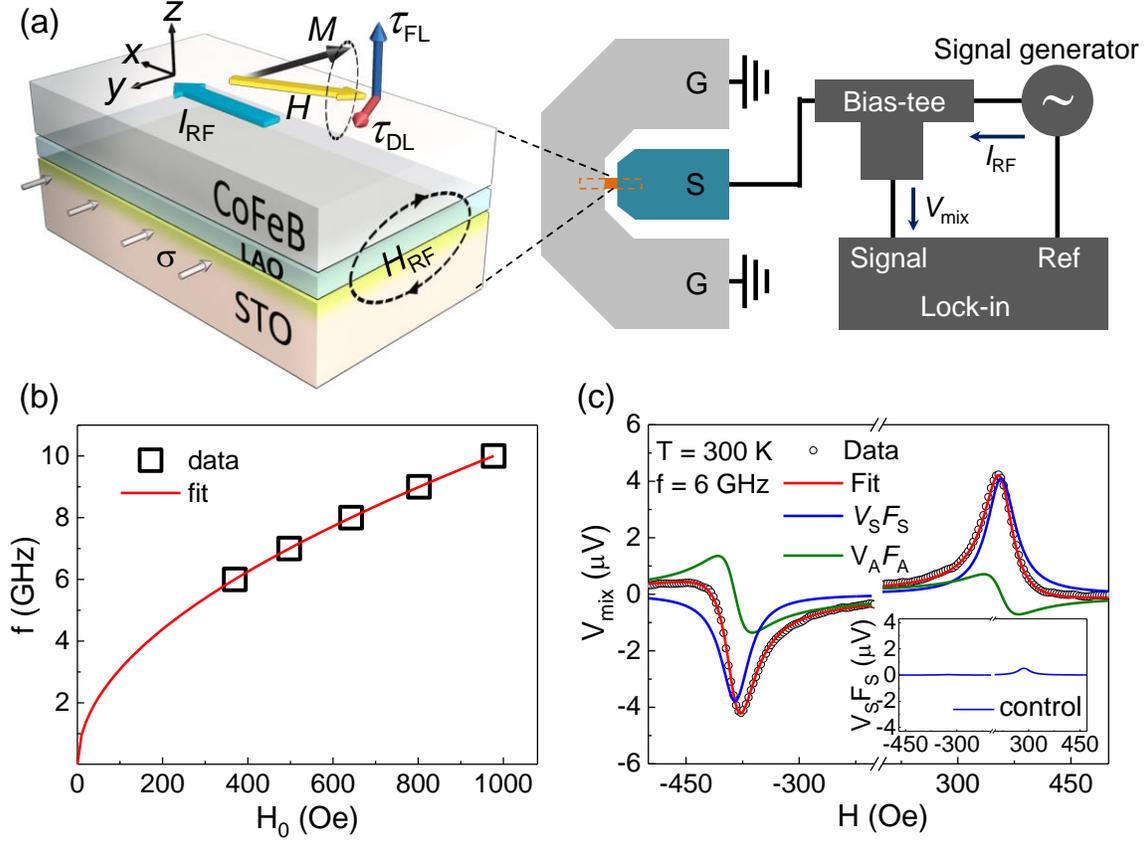

**Figure 2.** (a) The schematic of the ST-FMR measurement setup and illustration of the ST-FMR device with the SOT induced magnetization dynamics. (b) Resonance frequency $f$ against the resonant field $H_0$ for a STO/LAO (8 uc)/CFB (7 nm) device with a fit. (c) A representative ST-FMR signal (open symbols) from a STO/LAO (8 uc)/CFB (7 nm) device at 6 GHz and 300 K with fits of whole signal (red lines), the symmetric Lorentzian ($V_S F_S$) component (blue lines) and the antisymmetric Lorentzian ($V_A F_A$) components (green lines). The inset shows only the symmetric component $V_S F_S$ from a control device with a single CFB layer.

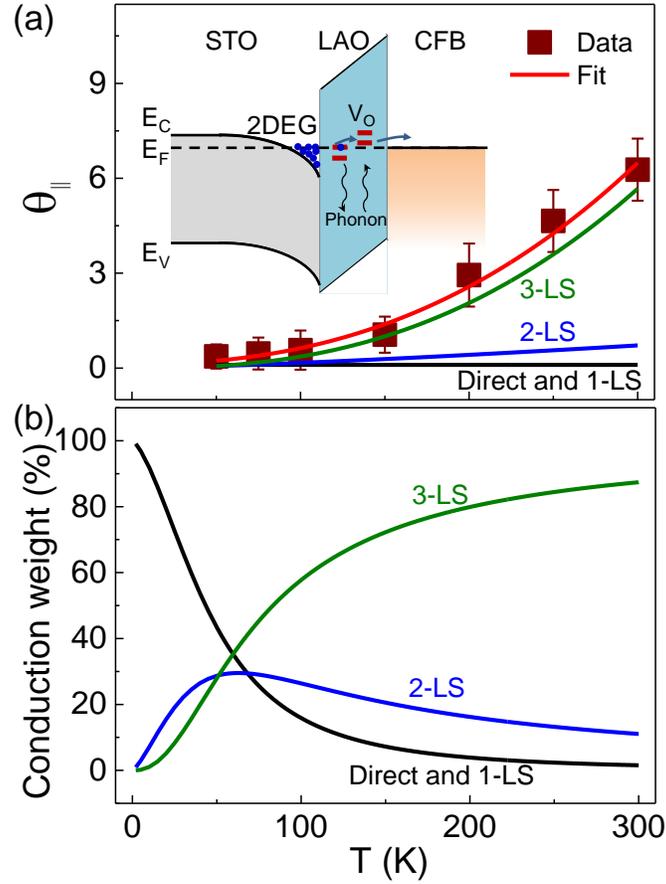

**Figure 3.** (a) SOT efficiency, $\theta_\parallel$, as a function of temperature and the fit (red curve). Each $\theta_\parallel$ represents the averaged value from three measurement frequencies (6–8 GHz), and the error bars are the standard deviation. The inset is the schematic of the spin-polarized electron inelastic tunneling process via the LS, such as oxygen vacancies, $V_O$. (b) The conduction weighted contributions of the respective direct elastic tunneling & 1-LS, 2-LS and 3-LS chains.

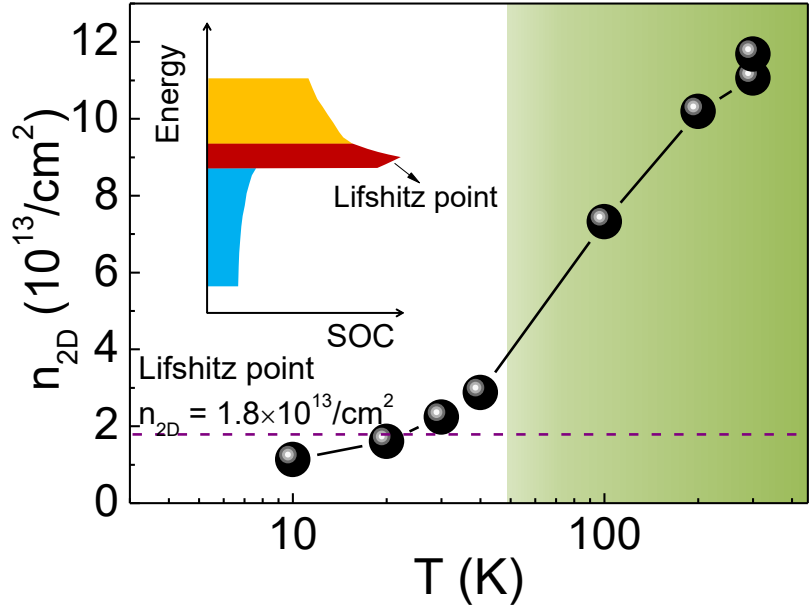

**Figure 4.** Temperature dependent sheet carrier concentration, $n_{2D}$, of the 2DEG layer at the STO/LAO interface. The Lifshitz point is denoted around the carrier concentration of ~$1.8\times10^{13}$/cm$^2$. The green shaded area represents the temperature range of our ST-FMR measurements. The inset shows the sketch of the the relationship between the strength of SOC and Fermi energy adapted from Ref. 6.